\documentclass[final]{article}

\PassOptionsToPackage{numbers, compress}{natbib}
 \usepackage[nonatbib]{nips_2018}

\usepackage[utf8]{inputenc} 
\usepackage[T1]{fontenc}    
\usepackage{hyperref}       
\usepackage{url}            
\usepackage{booktabs}       
\usepackage{amsfonts}       
\usepackage{nicefrac}       
\usepackage{microtype}      
\usepackage{cite, graphicx, subfigure,url,setspace}
\usepackage{algorithm}
\usepackage{algpseudocode,algorithm,algorithmicx}
\usepackage{amssymb, color}
\usepackage{bm}
\usepackage{amsmath}
\usepackage{CJK}

\newtheorem{corollary}{Corollary}
\newtheorem{assumption}{Assumption}
\newtheorem{theorem}{Theorem}

\title{A Decoupled Data Based Approach to Stochastic Optimal Control Problems}

\author{
   Dan Yu\\
 College of Astronautics\\
 Nanjing University of Aeronautics and Astronautics\\
 Nanjing, China, 210016 \\
  \texttt{yudan198811@hotmail.com} \\
   \And
  Mohammandhussen Rafieisakhaei\\
  Department of Electrical and Computer Engineering\\
  Texas A$\&$ M University\\
  College Station, TX, 77840 \\  
  \texttt{mrafieis@tamu.edu} \\
  \AND
  Suman Chakravorty\\
  Department of Aerospace Engineering\\
  Texas A$\&$ M University\\
  College Station, TX, 77840 \\  
  \texttt{schakrav@tamu.edu} \\
}

\begin{document}

\maketitle

\begin{abstract}
This paper studies the stochastic optimal control problem for systems with unknown dynamics. A novel decoupled data based control (D2C) approach is proposed, which solves the problem in a decoupled ``open loop-closed loop" fashion that is shown to be near-optimal. First, an open-loop deterministic trajectory optimization problem is solved using a black-box simulation model of the dynamical system using a standard nonlinear programming (NLP) solver. Then a Linear Quadratic Regulator (LQR) controller is designed for the nominal trajectory-dependent linearized system which is  learned using input-output experimental data.  Computational examples are used to illustrate the performance of the proposed approach with three benchmark problems.
\end{abstract}

\section{Introduction}\label{Section 1}
Stochastic optimal control problems, also known as Markov decision problems (MDPs), have found numerous applications in the Sciences and Engineering. In general, the goal is  to control a stochastic system subject to transition uncertainty in the state dynamics so as to minimize the expected running cost of the system.  In this work, we propose a novel data based approach to the solution of MDPs residing in continuous state and control spaces. Our approach proposes a rigorous decoupling of the open loop (planning) problem from the closed loop (feedback control) problem in the sense that the decoupled design is near optimal to the third order. Furthermore, this decoupling allows us to propose a highly data efficient approach to solving MDPs in a completely data based fashion (see Table \ref{comp_time} and \ref{ctime}). The approach proceeds in two steps:
\begin{itemize}
\item First, we optimize the nominal open loop trajectory of the system using a blackbox simulation model and an NLP solver.
\item Second, we identify the linear system governing perturbations from the nominal trajectory using random input-output perturbation data, and design an LQR controller for the linearized system.
\end{itemize}

\section{Related Work}
It is well known that the global optimal solution for MDPs can be found by solving the Hamilton-Jacobi-Bellman (HJB) equation \cite{dp_bertsekas}. The solution techniques can be further divided into model based and model free techniques, according as whether the solution methodology uses an analytical model of the system or it uses a black box simulation model, or actual experiments.

\textit{Model based Techniques:} In model based techniques, many methods \cite{dp_num} rely on a discretization of the underlying state and action space, and hence, run into the "curse of dimensionality (COD)", the fact that the computational complexity grows exponentially with the dimension of the state space of the problem.  The most computationally efficient among these techniques are trajectory-based methods such as  differential dynamic programming (DDP) \cite{ddp, sddp} which linearizes the dynamics and the cost-to-go function around a given nominal trajectory, and designs a local feedback controller using DP. The iterative Linear Quadratic Gaussian  (ILQG) \cite{ilqg1, ilqg2}, which is closely related to DDP,  considers the first order expansion of the dynamics (in DDP, a second order expansion is considered), and designs the feedback controller using Riccati-like equations, and is shown to be computationally more efficient. In both approaches, the control policy is executed to compute a new nominal trajectory, and the procedure is repeated until convergence. 

\textit{Model free RL Techniques:} In the model free solution of MDPs, the most popular approaches are the adaptive dynamic programming  (ADP) \cite{adp,adp_3} and  reinforcement learning (RL) paradigms \cite{rl_1, rl_3}. They are essentially the same in spirit, and seek to improve the control policy for a given black box system by repeated interactions with the environment, while observing the system's responses. The repeated interactions, or learning trials, allow these algorithms to construct a solution to the DP equation, in terms of the cost-to-go function,  in an online and recursive fashion. Another variant  of RL techniques is the so-called Q-learning method, and the basic idea in Q-learning is to estimate a real-valued function $Q(x, a)$ of states and actions instead of the cost-to-go function $V(x)$. For continuous state and control space problems, the cost-to-go functions and the Q-functions are usually represented in a functionally parameterized form, for instance,  in the linearly parametrized form $Q(x,a) = \theta' \phi(x,a)$,  where $\theta$ is the unknown parameter vector, and $\phi$ is  a pre-defined basis function, $(\cdot)'$ denotes the transpose of $(\cdot)$. Multi-layer (deep) neural networks may also be used as nonlinearly parameterized approximators instead of the linear architecture above. The ultimate goal of these techniques is the estimation/ learning of the parameters $\theta$ from learning trials/ repeated simulations of the underlying system. However, the size of the parameter $\theta$ grows exponentially in the size of the state space of the problem without a compact parametrization of the cost-to-go or Q function in terms of the a priori chosen basis functions for the approximation, and hence, these techniques are typically subject to the curse of dimensionality. Albeit a compact parametrization may exist, a priori, it is usually never known.  Recent work on "Deep" RL has shown promise to scale to continuous action and state space robotic learning problems \cite{DPG,DDPG,D4PG}, nonetheless, the amount of training required still seems prohibitive.
In the past several years, techniques based on the differential dynamic programming/ ILQG approach such as the RL techniques \cite{RLHD4, RLHD5, RLHD1} have shown the potential for RL algorithms to scale to continuous high dimensional robotic task planning and learning problems. 
For continuous state and control space problems, the method of choice is to wrap an LQR feedback policy around a nominal trajectory and then perform a recursive optimization of the feedback law, along with the underlying trajectory, via repeated simulations/ iterations. However, the parametrization can still be very large and can lead to the so-called ``policy chatter" phenomenon \cite{RLHD1}. 

\textit{Fundamentally, rather than solve the derived ``Dynamic Programming" problem as in the majority of the approaches above  that requires the optimization of the feedback law, our approach is to directly solve the original stochastic optimization problem in a ``decoupled open loop -closed loop" fashion wherein:  1) we solve an open loop deterministic optimization problem to obtain an optimal nominal trajectory in a model free fashion, and then 2) we design a closed loop controller for the resulting linearized time-varying system around the optimal nominal trajectory, again in a model free fashion. Nonetheless, the above ``divide and conquer" strategy can be shown to be near optimal to the third order. }  
The primary contributions of the proposed approach are as follows:
1) it shows a near optimal parametrization of the feedback policy in terms of an open loop control sequence, and a linear feedback control law, 2) it shows rigorously that the open loop and closed loop learning can be decoupled, which 3) results in the D2C algorithm that is highly data efficient when compared to state of the art RL and model-based techniques (Table \ref{comp_time} and \ref{ctime}). 


The rest of the paper is organized as follows. In Section \ref{Section 2}, the basic problem formulation is outlined.In Section \ref{Sec_theorem}, a decoupling result which solves the MDP in a ``decoupled open loop-closed loop " fashion is briefly summarized. In Section \ref{Section 3}, we propose a decoupled data based control algorithm, with discussions of implementation problems. In Section \ref{Section 4}, we  test the proposed approach using three typical benchmarking examples with comparisons to the model based iLQG approach and state of the art RL techniques.

\section{Problem Setup}\label{Section 2}
Consider the following discrete time nonlinear dynamical system:
\begin{align} \label{original system}
x_{k + 1} = f(x_k) + B_k( u_k + \epsilon  w_k), 
\end{align}
where $x_k \in \mathbb{R}^{n_x}$,   $u_k \in \mathbb{R}^{n_u}$ are the state, measurement and control vector at time $k$, respectively.  The process noise $w_k$ is assumed as zero-mean, uncorrelated Gaussian white noise, with covariance $W$, and  $\epsilon$ is a noise scaling parameter.


We consider the following stochastic optimal control problem. 

\paragraph{Stochastic Control Problem:} For the system with unknown nonlinear dynamics $f(\cdot)$, the optimal control problem is to find the control policies $\pi = \{\pi_0, \pi_1, \cdots, \pi_{N -1} \}$ in a finite time horizon $[0, N]$, where $\pi_k$ is the control policy at time $k$, i.e., $u_k = \pi_k(x_k)$, such that for a given initial  state $x_0$,  the cost function
$
J_{\pi} = E(\sum_{k = 0}^{N-1} c(x_k, u_k) + \phi(x_N)), 
$
is minimized, where $\{ c(\cdot,\cdot) \}_{k = 0}^{N - 1}$ denotes the  incremental cost function, and $\phi(\cdot)$ denotes the terminal cost. The expectation is taken over all randomness. 

\section{A Decoupling Result}\label{Sec_theorem}
In the following, we summarize the key ``Decoupling" result at the basis of our data based feedback control design technique. The proofs are shown in Appendix A and Appendix B.

Let $x_{k+1}^{\epsilon} = f(x_k^{\epsilon}) + B_k(u_k + \epsilon w_k)$ denote the state evolution of the system. Let $\bar{x}_{k+1} = f(\bar{x}_k) + B_k\bar{u}_k$ denote the evolution of a nominal state trajectory $\bar{x}_k$, where $\bar{u}_k$ is the nominal control action. Let the incremental cost function have the form $c(x,u) = l(x) + \frac{1}{2} u'Ru$. Now, let us define a perturbation of the true trajectory from the nominal, $\delta x_k^{\epsilon} = x_k^{\epsilon} - \bar{x}_k$. The evolution of the perturbed state is given by the equation: $\delta x_{k+1}^{\epsilon} =A_k \delta x_k^{\epsilon} + B_k \delta u_k + \epsilon w_k + r_k(\delta x_k^{\epsilon})$, where $A_k = \frac{\partial f}{\partial x} |_{\bar{x}_k}$, $\delta u_k$ is a perturbation in the nominal control effort and $r_k(.)$ is a residual term containing the second and higher order dynamics. Let $J_k(x_k^{\epsilon})$ denote the optimal cost-to-go at time $k$ from state $x_k^{\epsilon}$. The optimal cost-to-go function may be expanded in terms of the perturbed state as $J_k(x_k^{\epsilon}) = \bar{J}_k + G_k \delta x_k^{\epsilon} + \delta x_k^{\epsilon'}P_k \delta x_k^{\epsilon} + q_k(\delta x_k^{\epsilon})$, where $q_k(.)$ denotes the third and higher order terms in the cost-to-go function. Also, we can expand the incremental cost function $l(x_k) = \bar{l}_k + L_k\delta x_k^{\epsilon} + \delta x_k^{\epsilon'}L_{kk}\delta x_k^{\epsilon} + s_k(\delta x_k^{\epsilon})$.
\begin{assumption}\label{as1}
We assume that the functions $f(.)$, $J_k(.)$ and $l(.)$ are sufficiently smooth such that given any nominal trajectory $(\bar{x}_k, \bar{u}_k)$, they permit expansions in terms of the perturbed state $\delta x_k^{\epsilon}$ as above. Moreover, we assume that the residual functions $r_k(.), q_k(.)$ and $s_k(.)$ in the above expansions are all uniformly Lipschitz continuous in a neighborhood $N_k$ of the nominal trajectory $\bar{x}_k$, for all $k$.
\end{assumption}
\begin{theorem} \label{opt}
\textbf{Decoupling.} Let $\bar{x}_k, \bar{u}_k$ be an optimal nominal trajectory. Under Assumption \ref{as1}, the gain $G_k$ and covariance $P_k$ evolution equations of the cost-to-go function $J_k(.)$ are as follows:
\begin{align}
&G_k = L_k + G_{k+1} A_k, \label{OL} \\
&P_k = L_{kk} + \frac{1}{2} K_kR_kK_k' + A_k'P_{k+1}A_k - K_k'S_kK_k+ K_k'B_k'P_{k+1}B_kK_k + G_{k+1} \tilde{R}_{k,xx}, \label{feedback}
\end{align} 
where $G_N = \nabla \phi |_{\bar{x}_N}, P_N = \nabla^2 \phi |_{\bar{x}_N}$ denote the terminal conditions of the equations above, $\tilde{R}_{k,xx} = \nabla^2 f|_{\bar{x}_k}$ denotes the second order residual dynamics term, and $S_k = \frac{1}{2} R_k + B_k'P_{k+1}B_k$.
Further, let $\hat{J}_k(x_k^{\epsilon}) = \bar{J}_k + G_k \delta x_k^{\epsilon} + \delta x_k^{\epsilon'}P_k\delta x_k^{\epsilon}$.  Then, $|J_k(x_k^{\epsilon}) - \hat{J}(x_k^{\epsilon})|$ is an $O(||\delta x_k^{\epsilon}||^3)$ function.
\end{theorem}
\begin{theorem}\label{err_bound}
\textbf{Large Deviations bound.} Let Assumption \ref{as1} holds. Then, there exist finite constants $\alpha, \beta, \bar{K},$ independent of $\epsilon$, such that
\begin{align}
Prob(\max_{0\leq k \leq N} |\hat{J}_k(x_k^{\epsilon}) - J_k(x_k^{\epsilon})| \geq \bar{K} \gamma ^3 \epsilon^3) \leq \frac{\alpha}{\gamma}e^{-\beta \gamma^2}, \label{LD}
\end{align}
where $\gamma$ is some user-defined positive number.
\end{theorem}
A corollary of the above two results is the following.
\begin{corollary} \textbf{Global Optimaliy.}
Let $\bar{J}_t^1$ and $\bar{J}_t^2$ be the nominal costs for two different nominal trajectories with $\bar{J}_t^1 < \bar{J}_t^2$ for all $t$. Then, there exists a sufficiently small $\epsilon$, such that $\hat{J}_t^1(x_t^{\epsilon,1}) < \hat{J}_t^2(x_t^{\epsilon,2})$, given any arbitrarily high probability $1-\delta$.
\end{corollary}
Thus, if the nominal plan is a global optimum plan, then the feedback plan consisting of the nominal, and the linear feedback plan associated with it, is also globally optimal. 

\textit{Decoupling.} Due to the decoupling result, the $P_k$ equations which determine the feedback gains, do not affect the open loop ($G_k$) equations for an optimum nominal plan. Therefore, the nominal open loop design can be done completely independently of the feedback design. This can be done using any standard NLP solver in a blackbox fashion. Furthermore, the feedback design governed by the $P_k$ equations is a standard Ricatti equation, and can be solved given the $A_k, B_k, L_{kk}$ functions. However, these are rather straightforward to estimate given the nominal trajectory, using random rollouts of the perturbed optimal system. Therefore, the decoupling result breaks the feedback law design into two "simpler" decoupled problems of open loop and linear feedback design. Finally, but not the least, it suggests a near optimal (to the third order) parametrization of the feedback law: an open loop control sequence + a linear feedback law wrapped around it, $u_k(x_k^{\epsilon}) = \bar{u}_k + K_k\delta x_k^\epsilon$. The $G_k$ are the Lagrange multipliers/ co-states in the problem and Eq. \ref{OL} corresponds to the first order optimality conditions for the nominal control problem.

\textit{ILQG/DDP.} The condition in Eq. \ref{OL} is precisely when the iLQG/ DDP algorithms are deemed to have converged. However, that does not imply that the feedback gain in Eq. \ref{feedback} has converged. In fact, in iLQG/ DDP, once the open loop has converged, the Ricatti equation in Eq. \ref{feedback} would still need to be iterated in a policy iteration fashion till convergence to get the optimal feedback gain with respect to the optimized nominal trajectory. Again, this is evidence that the open loop and feedback problems are indeed decoupled.

\textit{Large Deviations.} The large deviations bound in Eq. \ref{LD} has two parameters $\gamma$ and $\epsilon$ to control the accuracy of the estimate and its probability of validity. Using a suitable $\gamma$, we can make the probability of violation (large deviation) as small as we desire. Similarly, by suitably choosing $\epsilon$, we can make the error bound as small as we desire. Thus, this result shows that the error in the approximation $|\hat{J}_k(.) - J_k(.)|$ incurred by only keeping the linear feedback term is $O(\epsilon^3)$ with an arbitrarily high probability.


\section{Decoupled Data Based Control (D2C) Design}\label{Section 3}

In this section, we propose a novel decoupled data based control (D2C) approach. First, we solve a noiseless open-loop optimization problem to find a nominal optimal trajectory and then we design a linearized closed-loop controller around the nominal trajectory, such that, with existence of stochastic perturbations, the state stays close to the optimal open-loop trajectory. The three-step framework to solve the stochastic feedback control problem may be summarized as follows. 
\begin{itemize}
\item Solve the open loop optimization problem using a general nonlinear programming (NLP) solver with a black box simulation model of the dynamics.
\item Linearize the system around the nominal open loop optimal trajectory, and identify the linearized time-varying system from input-output experiment data using a suitable system identification algorithm. 
\item Design an LQG controller which results in an optimal linear control policy around the nominal trajectory.
\end{itemize}
In the following section, we  discuss each of the above steps.

\subsection{Open Loop Trajectory Optimization}\label{sec_open}
Consider the noiseless nonlinear system:
\begin{align}
x_{k + 1} = f(x_k) + B_k u_k,\label{nominal}
\end{align}
with known initial state $x_0$. The open loop state optimization problem given an initial state $x_0$ is:
\begin{align}\label{cost_open}
\{ u_k^*\}_{k = 0}^{N - 1} = &\null  \arg \min_{\{u_k\}} \bar{J} (\{x_k\}_{k=0}^N, \{u_k\}_{k=0}^{N-1}), 
\end{align}
subject to the noiseless dynamics (Eq. \ref{nominal}).
The open loop optimization problem is solved using a general NLP solver, where the underlying dynamic model is used as a blackbox, and the necessary gradients and hessians are found by the solver typically using finite differencing, which is also highly amenable to parallellization. For example, open loop optimization using gradient descent \cite{gradient, sim_opt} is summarized  in Appendix C.

\subsection{Linear Time-Varying System Identification}\label{sec_tvera}
Denote the optimal open-loop control  as $\{\bar{u}_k \}_{k = 0}^{N - 1}$, and the corresponding nominal state as $\{\bar{x}_k\}_{k = 0}^{N-1}$.  We linearize the system (\ref{original system}) around the nominal trajectory  $\{\bar{x}_k\}$, assuming that the control and disturbance enter through the same channels and the noise is purely additive (these assumptions are only for simplicity and can be relaxed easily): 
\begin{align}\label{perturbation system}
& \delta x_{k+ 1} = A_k \delta x_k + B_k (\delta u_k + w_k), 
\end{align}
where $\delta x_k = x_k  - \bar{x}_k$ describes the state deviations from the nominal  trajectory, $\delta u_k = u_k - \bar{u}_k$ describes the control deviations, 
and
$
 A_k = \frac{\partial f(x, u, w)}{\partial x}|_{\bar{x}_k, \bar{u}_k, 0}, B_k = \frac{\partial f(x, u, w)}{\partial u}|_{\bar{x}_k, \bar{u}_k, 0}.   
$

Consider system (\ref{perturbation system}) with zero noise and  $\delta x_0 = 0$,  the input-output relationship is given by:
$
\delta x_k = \sum_{j = 0}^{k - 1} h_{k, j} \delta u_j, 
$
where $h_{k, j}$ is a generalized Markov parameters, and is defined by:
$
h_{k, j} = A_{k - 1} A_{k - 2} \cdots A_{j+1} B_j$.\\
\textit{Partial Realization Problem \cite{antoulas, skelton1}:} Given a finite sequence of Markov parameters $h_{k, j} \in \Re^{n_x \times n_u}, k = 1, 2, \cdots, s,  j = 0, 1, \cdots, k$, the partial realization problem consists of finding a positive integer $n_r$ and LTV system $(\hat{A}_k, \hat{B}_k, \hat{C}_k)$, where $\hat{A}_k \in \Re^{n_r \times n_r}, \hat{B}_k \in \Re^{n_r \times n_u}, \hat{C}_k \in \Re^{n_x \times n_r}$, such that the identified generalized Markov parameters $\hat{h}_{k, j} \equiv  \hat{C}_k \hat{A}_{k - 1} \hat{A}_{k - 2} \cdots \hat{A}_{j+1} \hat{B}_j= h_{k, j}$. Then $(\hat{A}_k, \hat{B}_k, \hat{C}_k)$ is called a partial realization of the sequence $h_{k, j}$. 

We solve the partial realization problem using the time-varying ERA, and construct the identified deviation system
\begin{align}\label{rom}
\delta a_{k  +1} = \hat{A}_k \delta a_k + \hat{B}_k (\delta u_k +  w_k), 
\;\delta x_k = \hat{C}_k \delta a_k, 
\end{align}
where $\delta a_k \in \Re^{n_r}$ denotes the reduced order model (ROM) deviation states. Time-varying ERA starts by estimating the generalized Markov parameters via least squares, using input-output experiments consisting of random rollouts of the optimized nominal system, i.e., by simulating the system under control actions $u_k = \bar{u}_k + \omega_k$, where $\bar{u}_k$ is the nominal control action at time $k$ and $\omega_k$ is a random noise perturbation. Then, it constructs a generalized Hankel matrix, and solves the singular value decomposition (SVD) problem of the constructed Hankel matrix. The dimension $n_r$ of the ROM is such that $n_r << n_x$ when $n_x$ is large, where $n_x$ is the dimension of the state, thereby automatically providing a compact parametrization of the problem. The details of the time-varying ERA can be found in \cite{tv_era}, and is briefly summarized in Appendix D.


\subsection{Feedback Controller Design}
Given the identified deviation system (\ref{rom}), we design the closed-loop controller to  follow the optimal nominal trajectory, which is to minimize the quadratized cost function
$
J_f = \sum_{k = 0}^{N - 1} (\delta \hat{a}_k' Q_k \delta \hat{a}_k+ \delta u_k' R_k \delta u_k ) + \delta \hat{a}_N' Q_N \delta \hat{a}_N,
$
where $\delta \hat{a}_k$ denotes the estimates of the deviation state $\delta a_k$, $Q_k, Q_N$ are positive definite, and $R_k$ is positive semi-definite. For the linear system (\ref{rom}), the ``separation principle" of linear control theory (not the Decoupling result of Section \ref{Sec_theorem}) can be used \cite{dp_bertsekas}. Using this result, the design of the optimal linear stochastic controller can be separated into the independent design of an optimal Kalman filter and a fully observed optimal LQR controller.  The details of the design is standard \cite{dp_bertsekas} and is omitted here. 

The Decoupled Data Based Control Algorithm is summarized in Algorithm \ref{algo_sc}.

\begin{algorithm}[!htbp]
\caption{Decoupled Data Based Control Algorithm}
\label{algo_sc}
 \begin{algorithmic}[1]
 \State Solve the deterministic open-loop optimization problem for optimal open loop nominal trajectory $(\{\bar{u}_k\}_{k = 0}^{N-1}, \{\bar{x}_k\}_{k = 0}^N)$ using a general NLP solver (Section \ref{sec_open}).  
 \State Identify the LTV system $(\hat{A}_k, \hat{B}_k, \hat{C}_k)$ via time-varying ERA (Section \ref{sec_tvera}).
 \State Solve the decoupled Riccati equations  using LTV system for feedback gain $\{ L_k \}_{k = 0}^{N}$. 
 \State Set $k = 0$, given initial estimates $\delta \hat{a}_0 = 0, P_0$.
\While{$k \leq N - 1$}
 \State \begin{align}
& u_k = \bar{u}_k - L_k \delta \hat{a}_k, \nonumber \\
& x_{k + 1} = f(x_k, u_k , w_k), \nonumber \\
\end{align}
Update $\delta \hat{a}_k$ using Kalman Filter.  
\State $k = k + 1$.
 \EndWhile
\end{algorithmic}
\end{algorithm}

\subsection{Discussion}\label{sec_dis}

\textit{Complexity:} The model free open loop optimization problem has complexity $O(n_u)$, where $n_u$ is the number of inputs, the LTV system identification step is again $O(n_u n_y)$, and the LQG feedback design has complexity $O(n_r^2)$, where $n_r$ is the order of the ROM from the LTV system identification step. Suppose we were to use an ILQG based design such as in \cite{RLHD1, RLHD2}, the complexity of the controller/ policy parametrization is $O(n_u n_x^2)$. Moreover, the policy evaluation step would require the estimation of a parameter of the size $O(n_x^4)$. For POMDPs,since $n_r<<n_x$ typically, the complexity of D2C approach is several orders of magnitude smaller.

\textit{Optimality:} The open loop law generated by the NLP solver can be guaranteed to be locally optimal under usual regularity conditions. Theorems \ref{opt}  and \ref{err_bound} show that the decoupled law is $O(\epsilon^3)$ optimal, and therefore, shows robust behavior even with moderate to high levels of noise (please see Section \ref{Section 4}). 

\section{Benchmark Examples}\label{Section 4}

In this section, we illustrate the D2C approach using three benchmark examples from the RL literature, with comparison to the model based iLQG approach. We also compare the data efficiency of the proposed approach with respect to reported results in the RL literature.

The noise-to-signal ratio (NSR) is defined as:
$NSR = \sqrt{\frac{W}{\|u\|}},$
where $W$ is the process noise covariance, and $u$ is the control input.

\begin{figure}[!htbp]
\centering
\subfigure[Cart-Pole]{
\includegraphics[width= 0.32\textwidth]{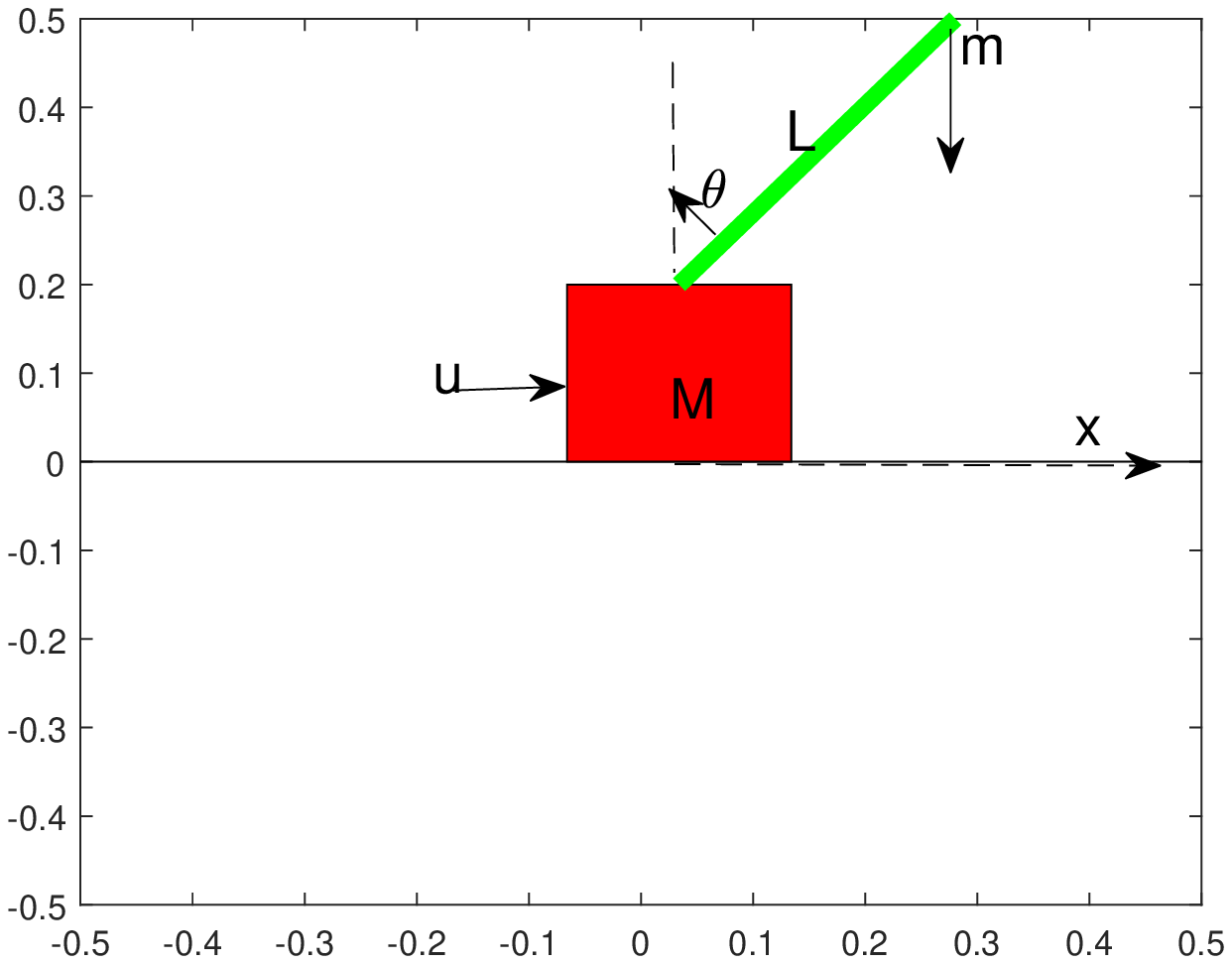}}
\subfigure[Cart-Two-Pole]{
\includegraphics[width= 0.32\textwidth]{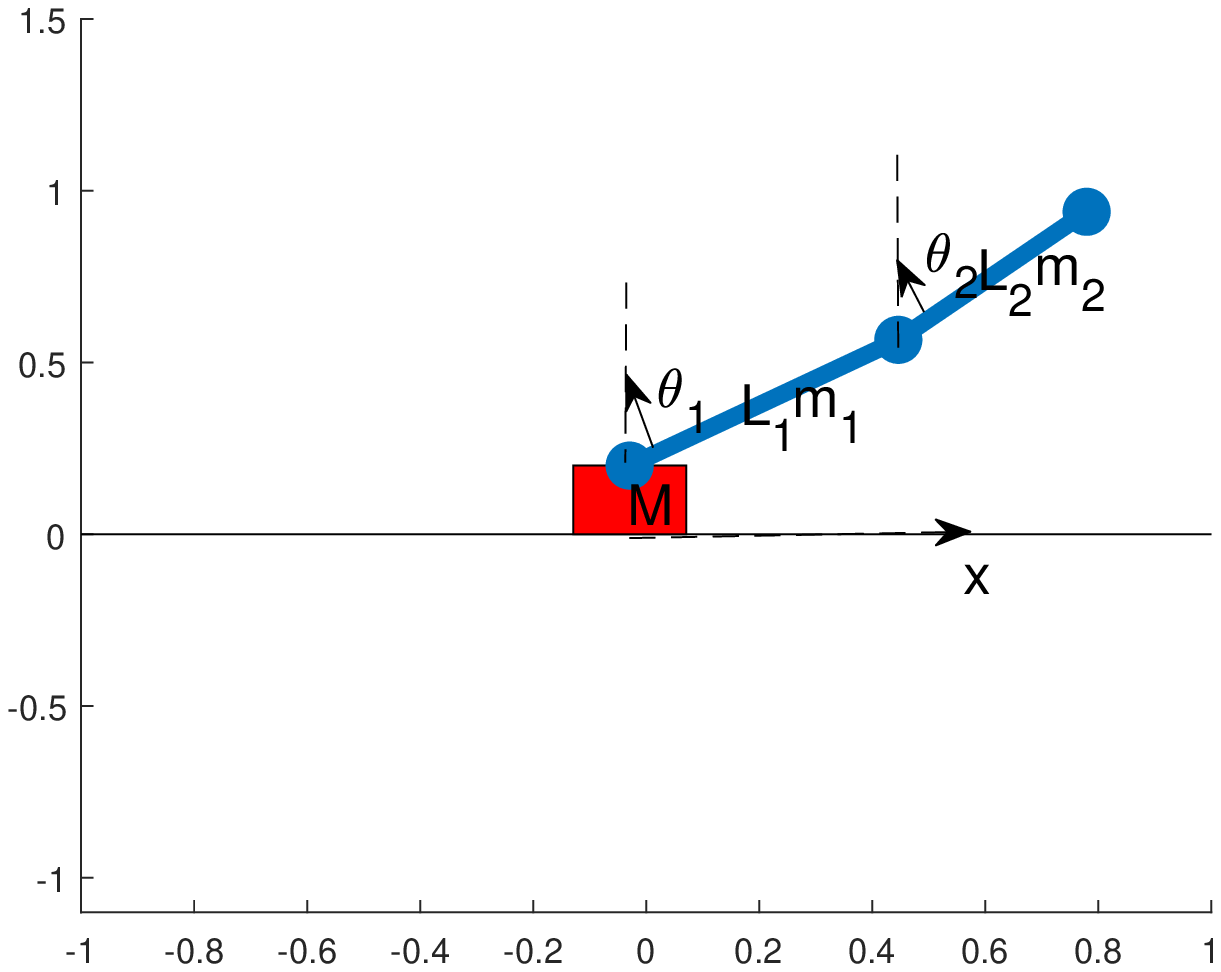}}
\subfigure[Acrobot]{
\includegraphics[width= 0.32\textwidth]{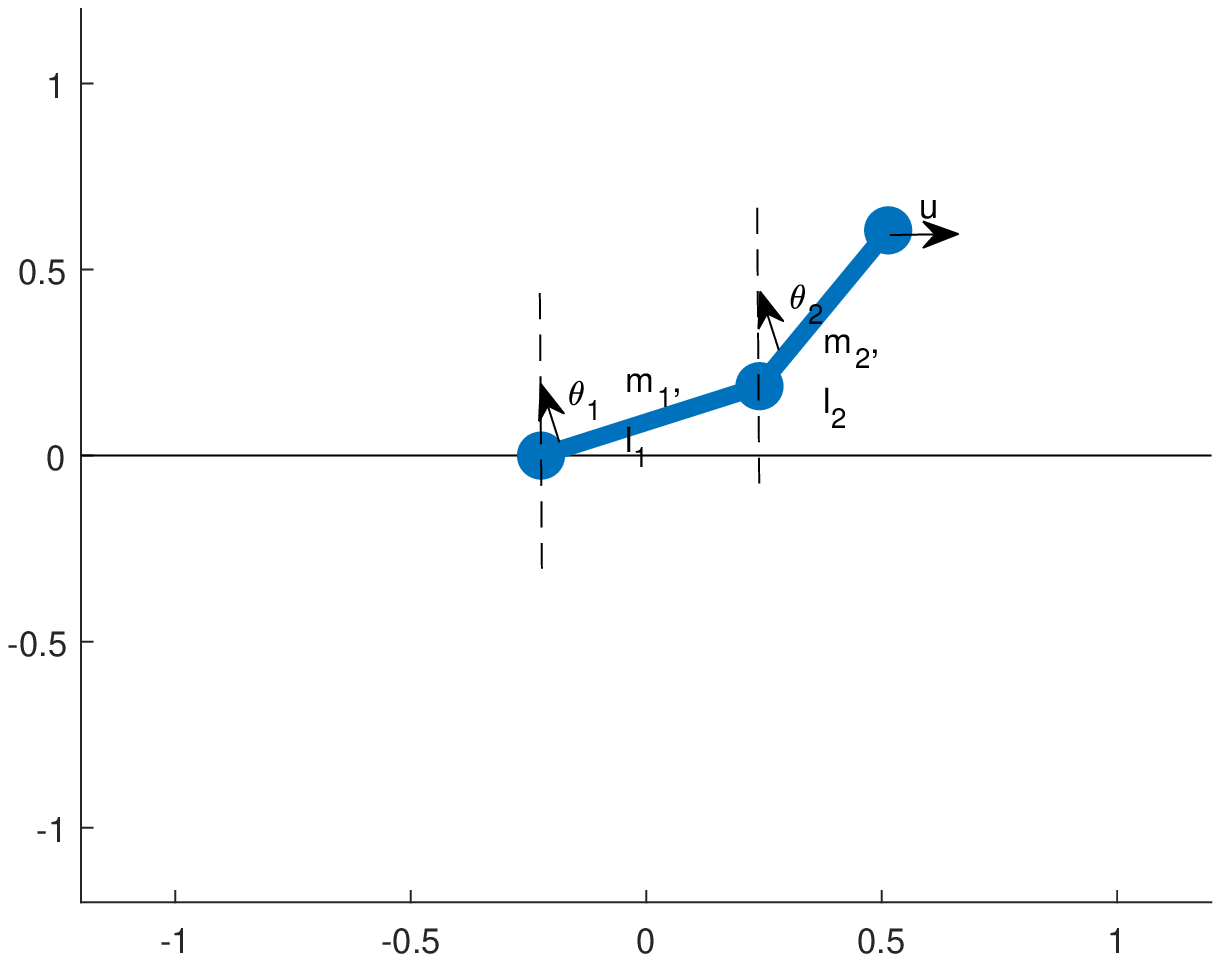}}
\caption{Benchmarking Example}\label{mod}
\end{figure}

We test the D2C approach using three benchmark fully observed examples: Cart-pole, Cart-two-pole (\cite{DDPG}) and Acrobot (\cite{ DeepRL_bench}).  The models  are shown in Fig. \ref{mod}. For Cart-pole and Cart-two-pole, the control is the force applied to the cart. For Acrobot, the torque is applied to the leg. For comparison, we use iLQG (\cite{ddp_code}), where we assume  that dynamic models of the examples required by iLQG are known (unknown to D2C).

\subsection{Cart-Pole} The control objective is to swing up within 3 seconds, and keep balance between $[3, 3.5]s$.  The initial states are $[x, \theta, \dot{x}, \dot{\theta}] = [0, \pi/4, 0, 0]$, and the target states are $[0, 0, 0, 0]$.  The parameters are $M= 1.0, m= 0.1, L= 0.5$. 

\subsection{Cart-Two-Pole} The control objective is to swing up within 3 seconds. The initial states are $[x,  \theta_1, \theta_2, \dot{x}, \dot{\theta}_1, \dot{\theta}_2] = [0, \pi/4, \pi/4, 0, 0, 0]$, and the target states are $[0, 0, 0, 0, 0, 0]$. The parameters are $M= 0.15, m_{1} = 0.6, m_{2} = 0.5, L_{1} = 0.6, L_{2} = 0.5$. 

\subsection{Acrobot} The control objective is to swing up within 5 seconds. The initial states are $[\theta_1, \theta_2, \dot{\theta}_1, \dot{\theta}_2] = [\pi/2, \pi/2, 0, 0]$, and the target states are $[0, 0, 0, 0]$. The parameters are $ m_{1} = m_{2} = 1, l_1 = l_2= 0.5, \mu_1 = \mu_2 = 0.05$, where $\mu_1,\mu_2$ are the friction coefficient of body and legs respectively. 

The averaged computational time using D2C approach is shown in Table \ref{comp_time}.  The computational complexity comparison between D2C and iLQG is shown in Table \ref{ctime}. The D2C open loop optimization problem is solved using Matlab NLP solver fmincon, with an interior-point algorithm.  The iLQG approach is implemented utilizing the Matlab toolbox provided in \cite{ddp_code}. 

\begin{table}[!htbp]
\caption{Averaged Computational Time (s)} \label{comp_time}
\centering
\begin{tabular}{c|c|c|c|c|c|}
&Open Loop & LTV ID& LQR & KF & Total Time\\
\hline
Cart-Pole & 22.5& 4.79 & 0.01 & 0.38 & 27.68\\
\hline
Cart-Two-Pole&52.69&  4.97 & 0.01 & 0.36 &57.67\\
\hline
Acrobot& 309.82& 8.26& 0.015&0.41 & 318.5\\
\hline
\end{tabular}
\end{table}

\begin{table*}[!htbp]
\caption{Computational Complexity Comparison} \label{ctime}
\centering
\begin{tabular}{c|c|c|c|c|c|c|}
&\multicolumn{2}{|c|}{Cart-Pole} & \multicolumn{2}{|c|}{Cart-Two-Pole} & \multicolumn{2}{|c|}{Acrobot}\\
\hline
&   D2C & iLQG & D2C & iLQG & D2C& iLQG\\
 \hline
$\#$ Iterations & 255 &582 &106&131& 160 & 148 \\
\hline
$\#$ Learning Trials & $2.34 \times 10^4$ & $\times$ & $3.35 \times 10^4$& $\times$& $1.6 \times 10^5$ & $\times$ \\
\hline
Off-line   Time    &       27.3(s)   & 318.25(s)&    57.67(s)        &  816(s)& 318.1(s)& 442.27(s) \\
\hline
On-line  Time     &        0.38(s)   & $\times$ &    0.36(s)         & $\times$ &0.41 (s) &$\times$  \\
\hline
Total Time     &       27.68(s)  & 318.25(s)&    58.03(s)         &816(s) & 318.5(s)  & 442.27(s)  \\
\hline                                             
 \end{tabular}                  
 \end{table*}

\begin{figure}[!htbp]
\centering
\subfigure[Cart-Pole Balance]{
\includegraphics[width= 0.32\textwidth]{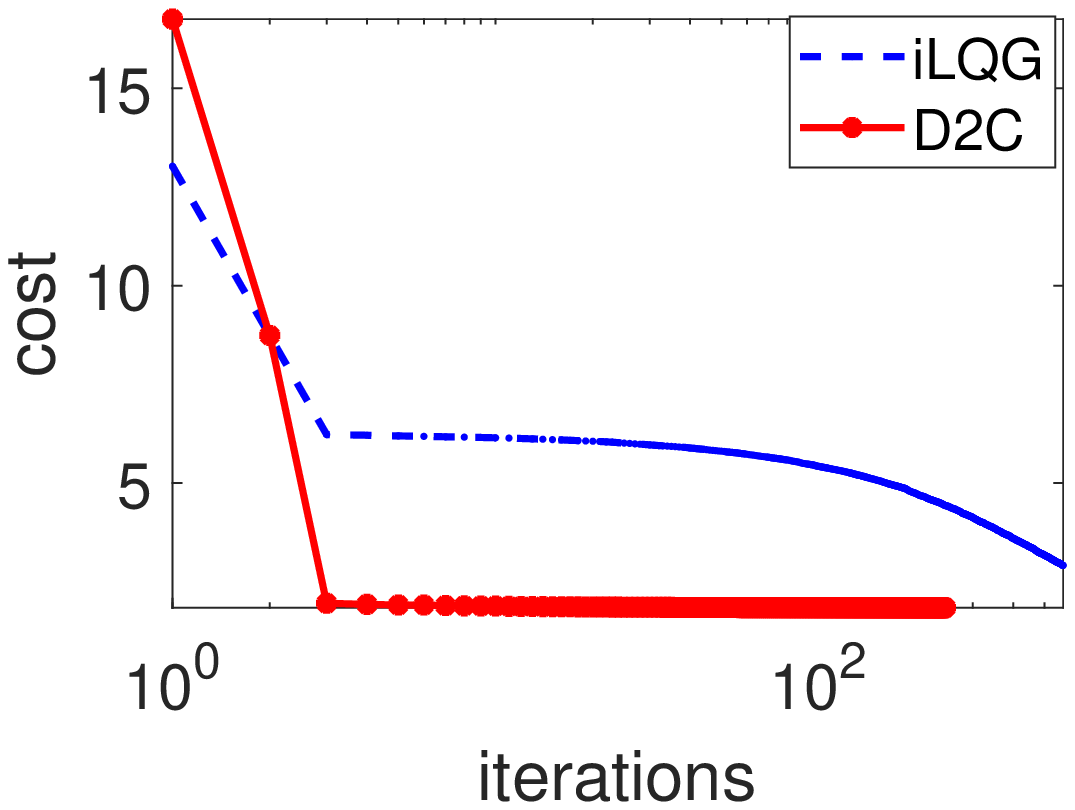}}
\subfigure[Cart-Two-Pole swing up]{
\includegraphics[width= 0.32\textwidth]{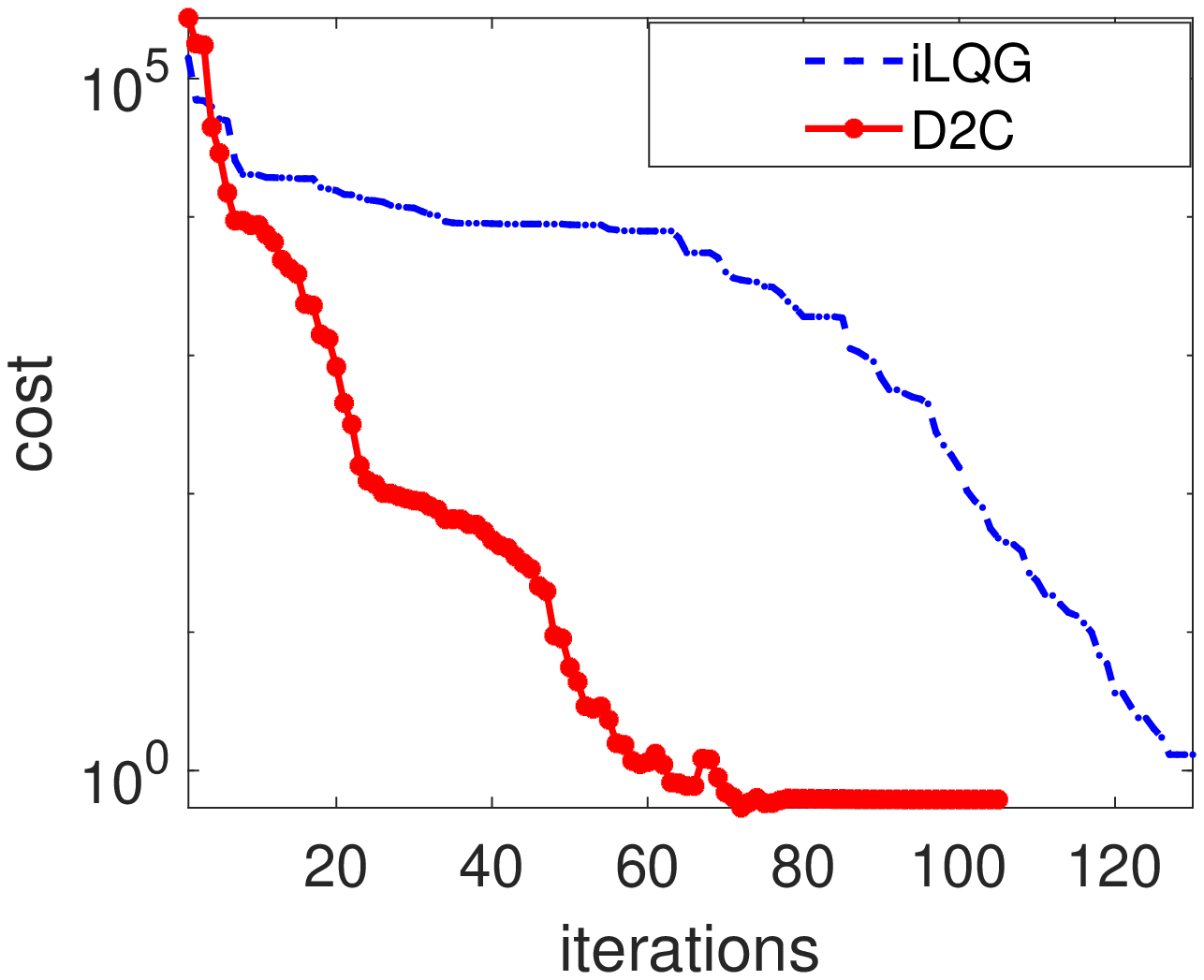}}
\subfigure[Acrobot swing up]{
\includegraphics[width= 0.32\textwidth]{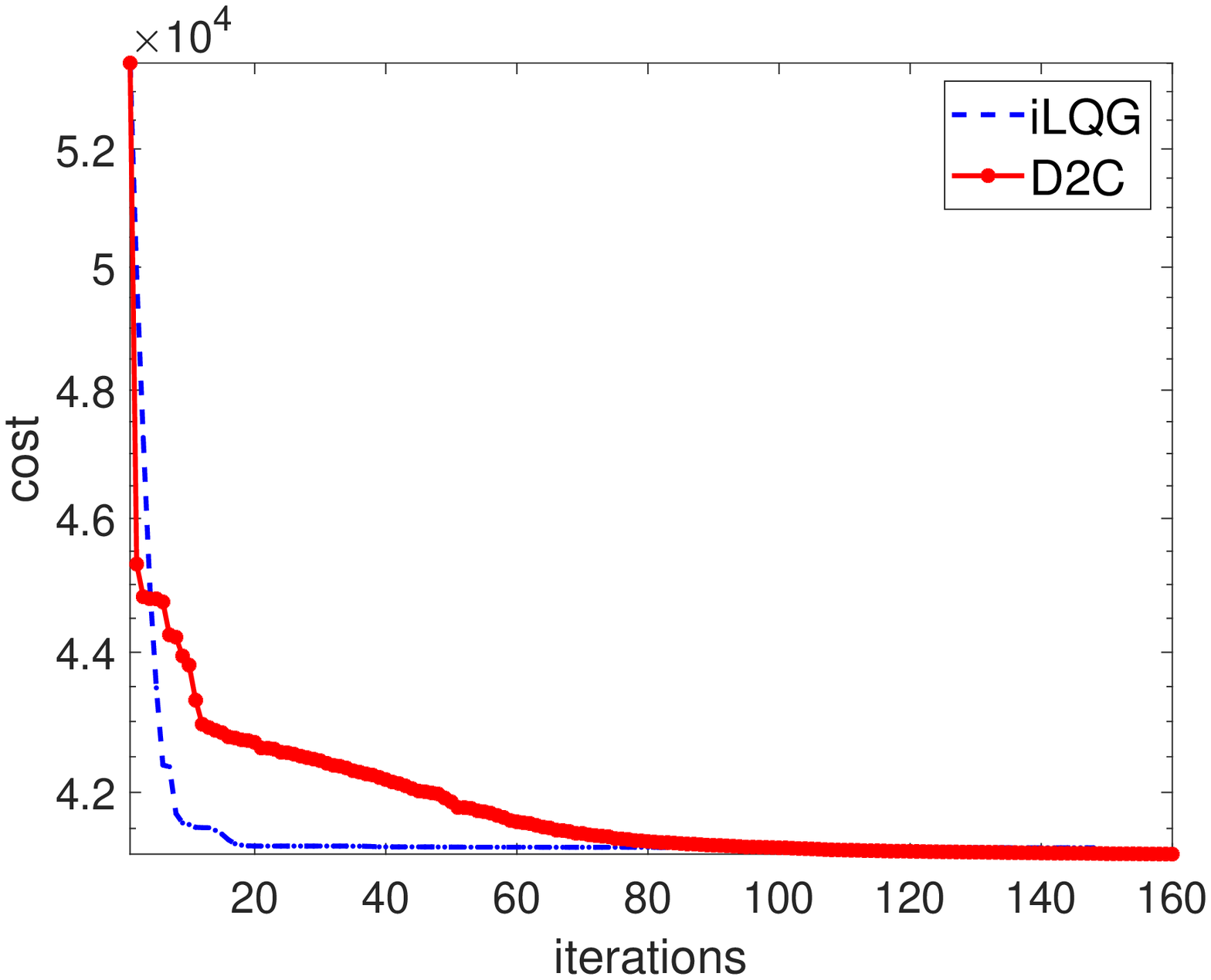}}
\caption{Comparison of the Convergence Rate.}\label{comp_cost}
\end{figure}

\begin{figure}[!htbp]
\centering
\subfigure[Cart-Pole]{
\includegraphics[width= 0.45\textwidth]{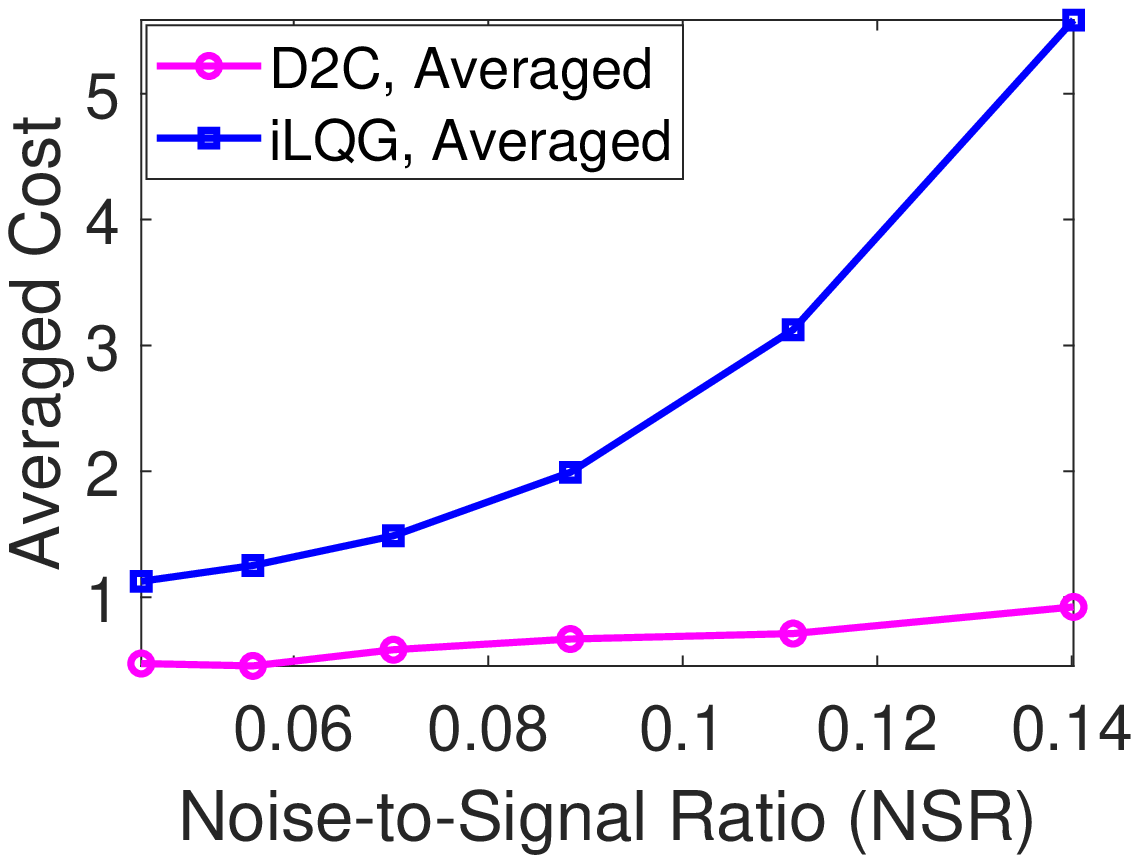}}
\subfigure[Cart-Two-Pole]{
\includegraphics[width= 0.45\textwidth]{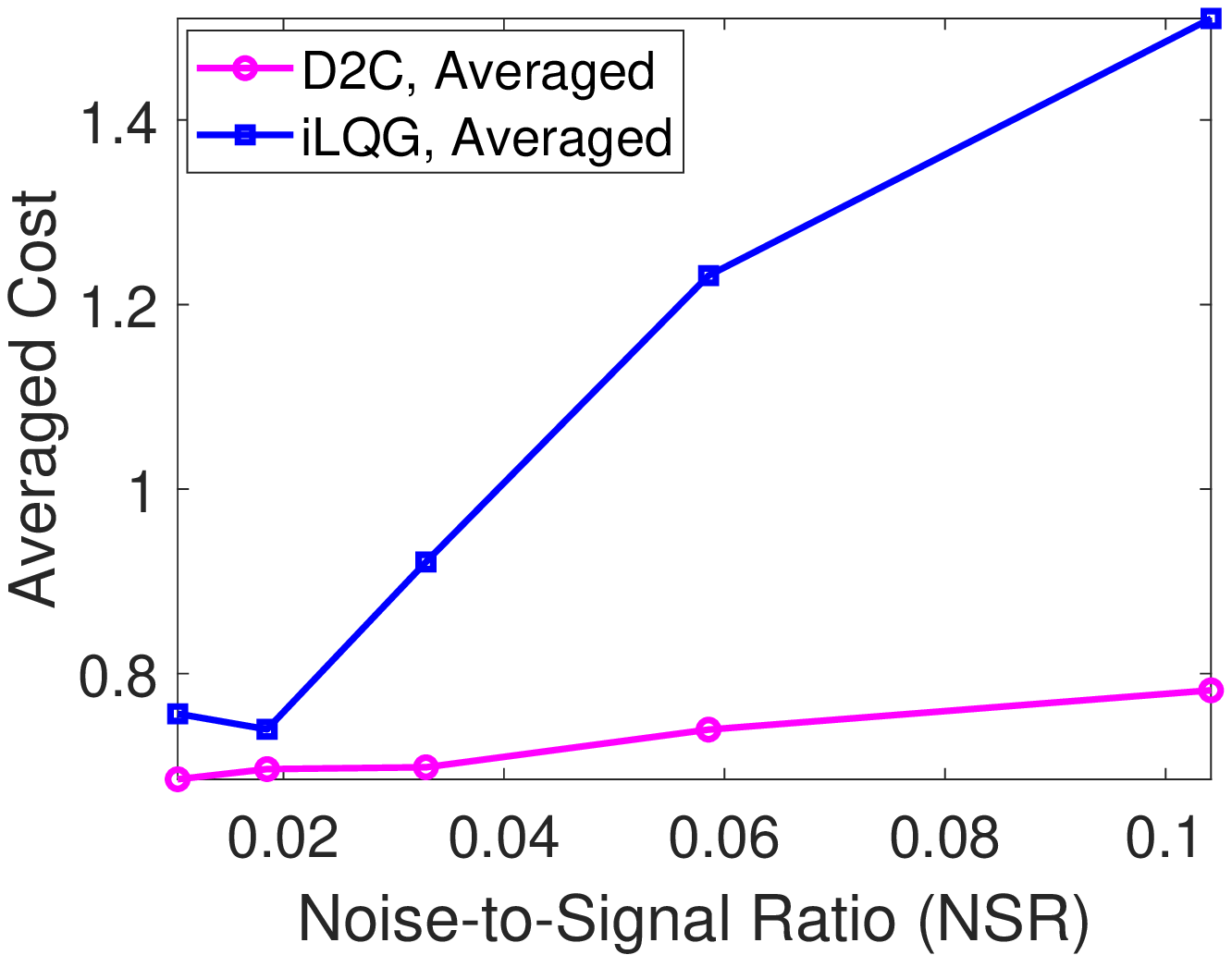}}
\caption{Comparison of the averaged cost.}\label{comp_nsr}
\end{figure}

\begin{figure}[!htbp]
\centering
\subfigure[$\theta_1$]{
\includegraphics[width= 0.23\textwidth]{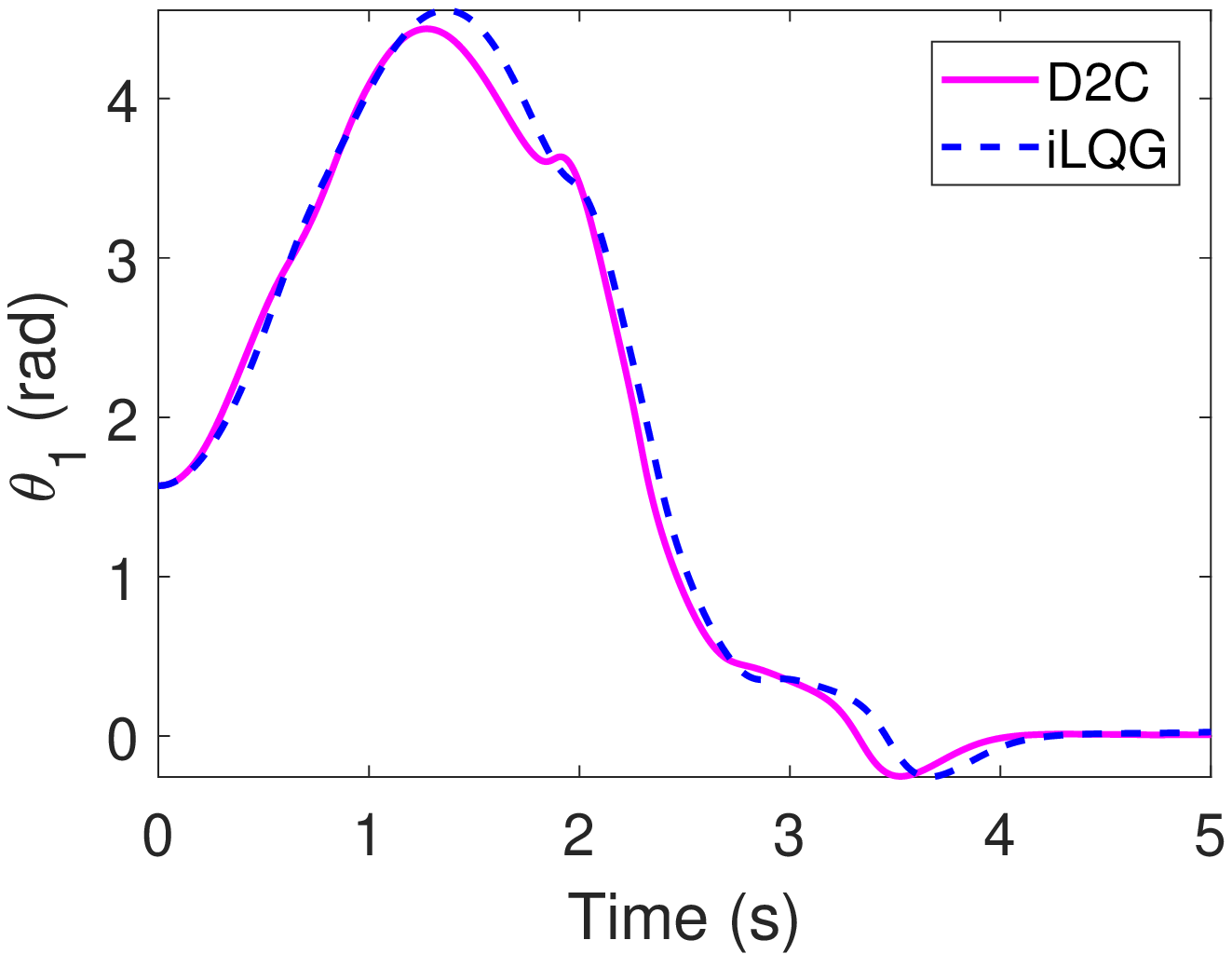}}
\subfigure[$\theta_2$]{
\includegraphics[width= 0.23\textwidth]{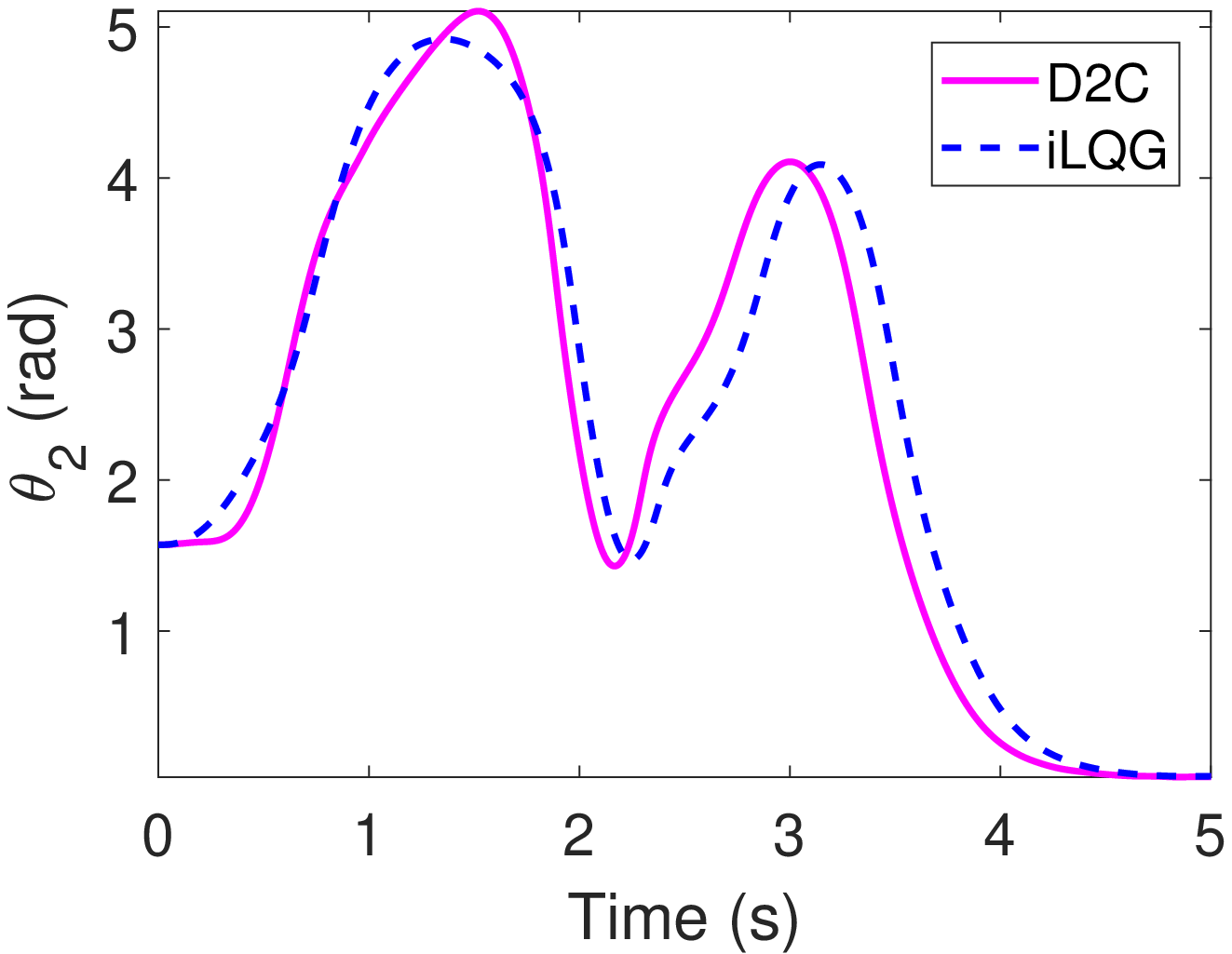}}
\subfigure[$\dot{\theta}_1$]{
\includegraphics[width= 0.23\textwidth]{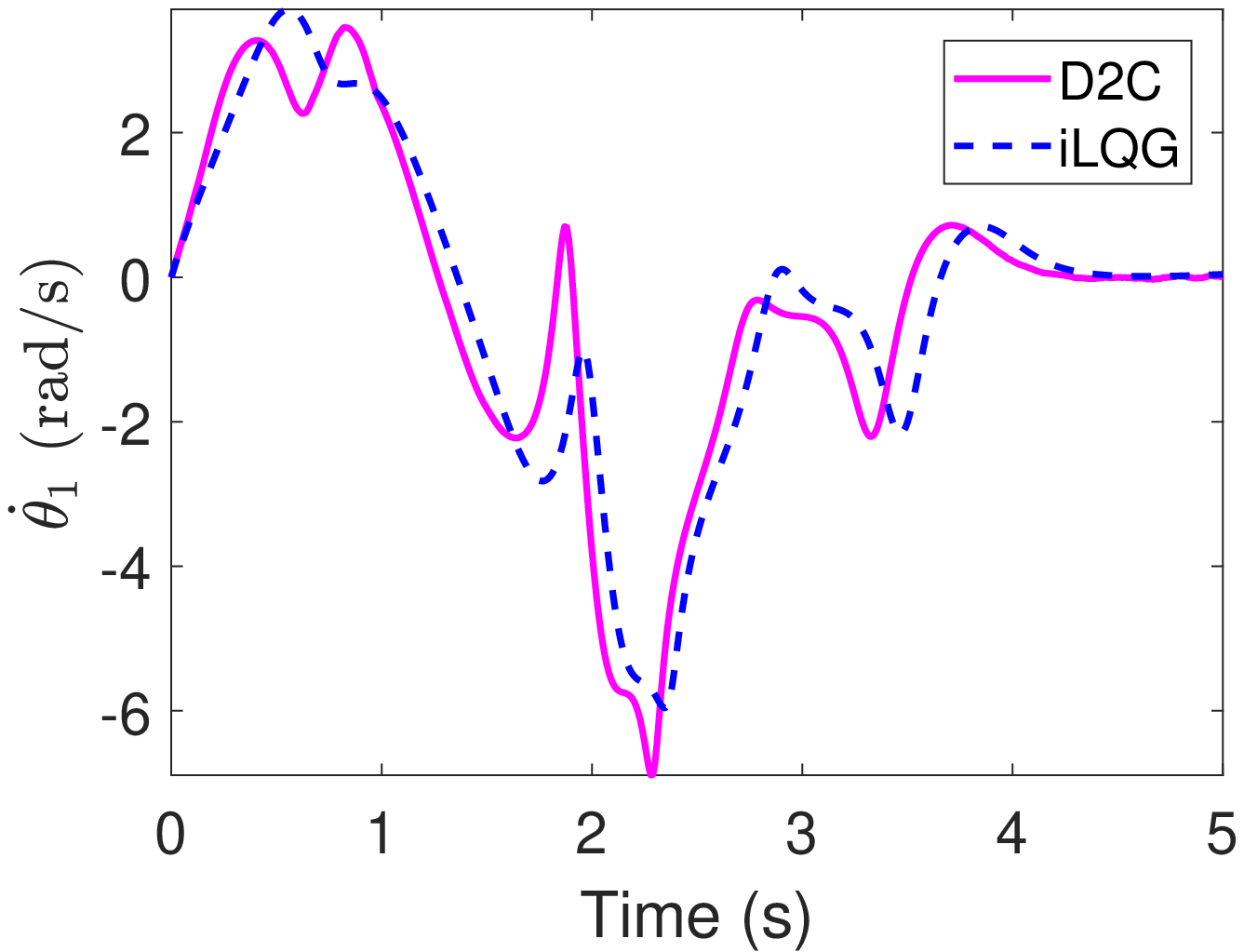}}
\subfigure[$\dot{\theta}_2$]{
\includegraphics[width= 0.23\textwidth]{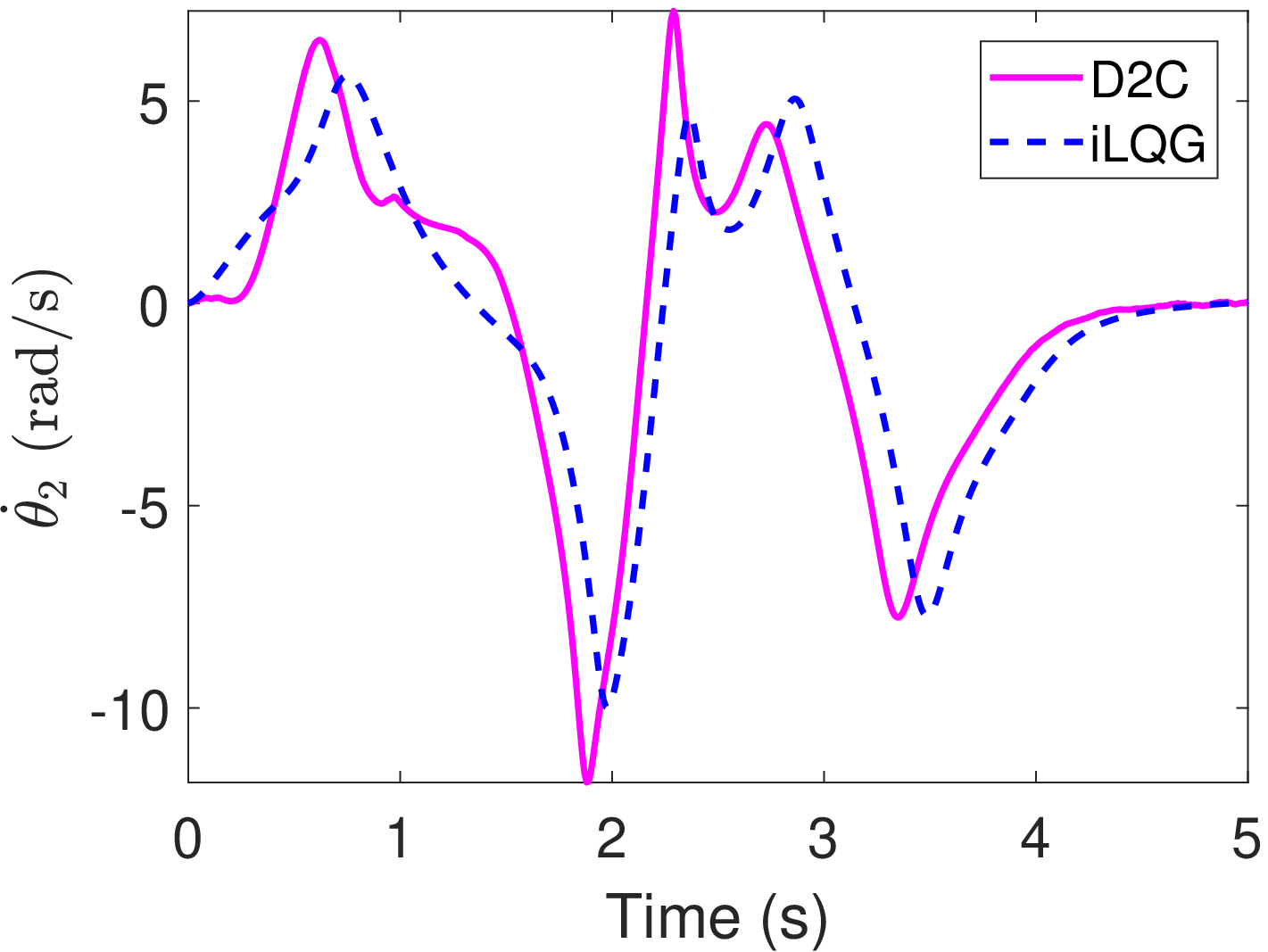}}
\caption{Performance Comparison for Acrobot}\label{acro_per}
\end{figure}

In all examples, the off-line computational time using D2C consists of the open loop optimization, LTV system identification and LQR design shown in Table \ref{comp_time}. The learning trials using D2C approach is the number of system rollouts used in the open loop optimization (22000, 32200 and $1.6 \times 10^5$) and the LTV system identification (1372,1324 and 565). Starting from the same initial guess, D2C and iLQG converge to the optimal solution with increasing number of iterations. In Fig. \ref{comp_cost}, the comparison of the convergence rate after the 1$^{st}$ iteration between D2C and iLQG for Cart-Pole balance and Cart-two-Pole swing up and Acrobot swing up are shown. The performance of the Cart-Pole swing up using both approaches are almost the same and are omitted here. We run 100 Monte-Carlo simulations, for Cart-pole and Cart-two-Pole, the comparison of the averaged cost using D2C and iLQG as a function of the NSR is shown in Fig. \ref{comp_nsr}. For Acrobot example, the averaged performance using D2C and iLQG when $NSR = 0.01$ are plotted in Fig \ref{acro_per}. In all examples, the same cost functions and initial guess are used, and are chosen ``fairly" such that both approaches could converge to a good solution.

It can be seen that 1) the computational time using D2C is much smaller than iLQG due to the decoupling as we do not have to solve the backward Ricatti equation at every iteration, (which takes up bulk of the time),  as required by iLQG.  The online computation using D2C is around $0.4s$; 2) the performance of D2C  approach is comparable to the iLQG approach. Further, the D2C approach seems less sensitive to the process noise when compared to the iLQG approach; and 3) we assume the dynamics is given to the  iLQG approach, which is unavailable to the D2C approach. 

\subsection{DeepRL comparison}
We note that for the tasks performed in this section, Deep RL algorithms such as the DDPG typically need $O(10^6)-O(10^8)$ learning trials, and several hours of learning time, even with parallelization. In contrast, we see that the D2C approach is able to accomplish the same tasks using $10^3-10^5$ learning trials, and in several minutes on a laptop computer. This, in our opinion, shows the data efficiency of the proposed D2C approach for continuous control problems \cite{DDPG, DeepRL_bench,D4PG}. However, this should not be surprising since the D2C approach is more efficient than a model based approach such as iLQG which is used as a benchmark by the Deep RL techniques \cite{DeepRL_bench}.

\section{Conclusion}
In this paper, we have proposed a decoupled data based design of the stochastic optimal control problem for systems with unknown nonlinear dynamics. First, we design a deterministic open-loop optimal trajectory.  Then we identify the nominal linearized system using time-varying ERA. The open-loop optimization and system identification are implemented offline, using the random perturbations/impulse responses of the system, and an LQG controller based on the ROM is implemented online.  We have tested the proposed approach on several fully observed benchmark examples, and showed the performance of the proposed approach with respect to the model based iLQG approach as well as state of the art RL techniques such as DDPG. Future work will generalize the proposed approach to large-scale partially observed systems.

\bibliographystyle{plain}
\bibliography{CDC_refs}
\end{document}



\maketitle
This supplementary material is organized as follows. We prove Theorem 1 and 2 in Appendices \ref{Proof_1} and \ref{Proof_2}. In Appendix \ref{sec_1}, we present the open loop optimization approach. In Appendix \ref{sec_2}, we review the time-varying eigensystem realization algorithm (ERA).
\appendix

\section{Proof of Theorem 1}\label{Proof_1}
Consider the Dynamic Programming equation corresponding to the true stochastic control problem: 
\begin{align}
J_k(x_k^{\epsilon}) = \min_{u_k} \{ c(x_k^{\epsilon}, u_k) + E [J_{k+1}(x_{k+1}^{\epsilon})]\}, \label{DP}
\end{align}
where the Q-function is given by:
\begin{align}
Q_k(x_k^{\epsilon}) = c(x_k^{\epsilon}, u_k) + E[J_{k+1}(x_{k+1}^{\epsilon})], \label{Q-fn}
\end{align}
and the terminal cost function is $J_N(x_N^{\epsilon}) = \phi_N(x_N^{\epsilon})$. Recall that $J_k(x_k^{\epsilon}) = \bar{J}_k + G_k\delta x_k^{\epsilon} + \delta x_k^{\epsilon'} P_k \delta x_k^{\epsilon} + q_k(\delta x_k^{\epsilon})$. Then, it follows that:
$J_{k+1}(x_{k+1}^{\epsilon}) = \bar{J}_{k+1} + G_{k+1} \delta x_{k+1} + \delta x_{k+1}'P_{k+1} \delta x_{k+1} + q_{k+1}$, where $\delta x_{k+1} = A_k \delta x_k^{\epsilon} + B_k \delta u_k + \epsilon B_k w_k + r_k$. Thus:
\begin{align}
&E[J_{k+1}(x_{k+1}^{\epsilon})] = \bar{J}_{k+1} + G_{k+1} A_k \delta x_k ^{\epsilon} + G_{k+1}B_k\delta u_k \nonumber\\
& + G_{k+1} r_k + \delta x_k^{\epsilon'} A_k'P_{k+1}A_k\delta x_k^{\epsilon} + \delta u_k'B_k'P_{k+1}A_k\delta x_k^{\epsilon} \nonumber\\
&+ \delta x_k^{\epsilon'}A_k'P_{k+1} B_k \delta u_k + \delta u_k'B_k'P_{k+1}B_k\delta u_N \nonumber\\
& + \epsilon^2 tr(P_{k+1}^{1/2}B_kB_k' P_{k+1}^{1/2})+ \delta u_k'B_k'P_{k+1} r_k + r_k'P_{k+1}B_k\delta u_k \nonumber\\
&+ \delta x_k'A_k'P_{k+1}r_k + r_k'P_{k+1}A_k\delta x_k + r_k' P_{k+1} r_k. \label{e1}
\end{align}
In a similar fashion, we can see that:
\begin{align}
c(x_k, u_k) =& \bar{l}_k + L_k \delta x_k^{\epsilon} + \delta x_k^{\epsilon'}L_{kk} \delta x_k^{\epsilon} + s_k
+ \frac{1}{2} \delta u_k' R_k \bar{u}_k\nonumber\\
& + \frac{1}{2}\bar{u}_k'R_k \delta u_k + \frac{1}{2}\delta u_k'R_k \delta u_k. \label{e2}
\end{align}
We may now use (\ref{e1}) and (\ref{e2}) to do the minimization of the Q-function on the RHS of  (\ref{DP}). Using the fact that $\min_{u_k} Q(x_k^{\epsilon}, u_k) = \min_{\delta u_k} Q(\delta x_k^{\epsilon}, \delta u_k)$, and setting $\frac{\partial Q_k}{\partial \delta u_k} = 0$, we obtain:
\begin{align}
\delta u_k^* =& \underbrace{-S_k^{-1}(R_k\bar{u}_k + G_{k+1}B_k)}_{\tilde{u}_k^*}  \underbrace{-S_k^{-1}(2B_k'P_{k+1}A_k)}_{K_k}\delta x_k^{\epsilon}\nonumber\\
& \underbrace{- S_k^{-1}(2B_k'P_{k+1}r_k)}_{p_k}, \label{e3}
\end{align}
where $S_k = \frac{1}{2} R_k + B_k'P_{k+1}B_k$.
If $\bar{u}_k$ is an optimal control sequence, then it follows from the first order optimality conditions that $R_k\bar{u}_k + G_{k+1}B_k=0$ for all $k$. Utilize this fact, substitute Eq. \ref{e3} back into  (\ref{Q-fn}) to get the optimal $Q_k(\delta x_k^{\epsilon}, \delta u_k^*)$, and then substitute the resulting $Q_k(\delta x_k^{\epsilon}, \delta u_k^*)$ back into the DP (\ref{DP}). Noting that $-\frac{1}{2} S_kK_k = B_k'P_{k+1}A_k$, and grouping the different powers of $\delta x_k^{\epsilon}$ on both sides of the resulting DP equation results in the following equations:
\begin{align}
G_k = &L_k + R_k\bar{u}_kK_k + G_{k+1}(A_k+B_kK_k), \nonumber\\
P_k = &L_{kk}+ \frac{1}{2}K_kR_kK_k' + A_k'P_{k+1}A_k - K_k'S_kK_k\nonumber\\
& + K_k'B_k'P_{k+1}B_kK_k + G_{k+1} \tilde{R}_{k,xx},
\end{align}
where $\tilde{R}_{k,xx} = \nabla^2 f|_{\bar{x}_k}$ is the second order terms in the residual dynamics term $r_k(\delta x_k^{\epsilon})$. 
Regrouping the $G_k$ equation, and noting that $\bar{u}_t$ is a nominally optimal sequence gives:
\begin{align}
G_k &= L_k + \underbrace{(R_ku_k + G_{k+1}B_k)}_{0}K_k + G_{k+1}A_k \nonumber\\
&= L_k + G_{k+1}A_k.
\end{align}
This completes the first part of our assertion. 

Noting that $|J_k(x_k^{\epsilon}) -\hat{J}_k(x_k^{\epsilon})| = |q_k (\delta x_k^{\epsilon})|$, it follows that the error incurred is $O(||\delta x_k^{\epsilon^3}||)$ since $q_k$ is the same. This completes the proof.

\section{Proof of Theorem 2}\label{Proof_2}
First, we state the following lemma for a large deviation bound on a linear Gaussian random process. The proof is relatively straightforward but tedious and omitted here.
\begin{lemma}
Consider the linear dynamical system $x_{k+1} = \bar{A}_k x_k + \epsilon B_k w_k$, where $w_k$ is a white noise process. There exist constants $\alpha, \beta$, only dependent on the system matrices $\bar{A}_k, B_k$ such that:
\begin{equation}
Prob(\max_{0\leq k\leq N} ||\delta x_k^l||>M) \leq \alpha \frac{\epsilon}{M} e^{-\beta \frac{M^2}{\epsilon^2}}. \nonumber
\end{equation}
\end{lemma}
Let $\delta x_k^l$ denote the solution of the fictitious linear system $\delta x_k^l = \bar{A}_k  \delta x_k^l + \epsilon B_k w_k,$ where $\bar{A}_k = A_k + B_kK_k$, and $K_k$ is the feedback gain matrix calculated from Theorem 1. 

\textbf{Proof of Theorem 2.}

Let $Prob(\max_{0\leq k \leq N} ||\delta x_k^l||\geq M) \leq \delta$. It can be shown that $||\delta x_k^{\epsilon} - \delta x_k^l||$ is an $O(\max _k ||\delta x^{\epsilon}_k||^2)$ function. Therefore, it follows that for sufficiently small $\delta x_k^{\epsilon}$: $Prob(\max_{0\leq k \leq N} ||\delta x_k^{\epsilon}||\geq M) \leq \delta$. Utilizing Assumption 1, we have that $|\hat{J}_k(\delta x_k^{\epsilon}) - J_k(\delta x_k^{\epsilon})| \leq L ||\delta x_k^{\epsilon}||^3,$ for sufficiently small $||\delta x_k^{\epsilon}||$. Thus, it follows that $Prob( \max_{0\leq k \leq N}|\bar{J}_k(\delta x_k^{\epsilon}) - J_k(\delta x_k^{\epsilon})| \geq L M^3) \leq \delta$. 

From Lemma 1, it follows that: $Prob(\max_{0 \leq k \leq N}||\delta x_k^l|| \geq M) \leq \alpha \frac{\epsilon}{M}e^{-\beta \frac{M^2}{\epsilon^2}}$, for some $\alpha, \beta $. Suppose now that $M = \gamma \epsilon$. This implies that $Prob(\max_{0 \leq k \leq N} ||\delta x_k ^l|| \geq \gamma \epsilon) \leq \frac{\alpha}{\gamma} e^{-\beta \gamma^2}$ which in turn implies that: $Prob(\max_{0\leq k \leq N} |\hat{J}_k(\delta x_k^{\epsilon}) - J_k(\delta x_k^{\epsilon})| \geq \bar{K} \gamma^3\epsilon^3)\leq \frac{\alpha}{\gamma} e^{-\beta \gamma^2}$, where $L= \bar{K}$. This completes the proof.

\textbf{Proof of Corollary 1.}
Let $i=1,2$. From Theorem 1 and 2, we have that, with high probability, $|J_t(x_t^{\epsilon,i}) - \hat{J}_t(x_t^{\epsilon,i})|$ is an $O(\epsilon^3)$ for all time $t$. Moreover, it can be seen that $|\bar{J}_t^i- J_t(x_t^{\epsilon,i})|$ is an $O(\epsilon^2)$ function with high probability. Therefore, it follows that $|\hat{J}_t(x_t^{\epsilon,i})  - \bar{J}_t^i|$ is also an $O(\epsilon^2)$ function with high probability. Therefore, choosing sufficiently small, we have that if $\bar{J}_t^1 < \bar{J}_t^2$ then it is also true that $\hat{J}_t(x_t^{\epsilon,1}) <\hat{J}_t^2(x_t^{\epsilon,2})$ for all time $t$.  

\section{Open Loop Optimization Algorithm}\label{sec_1}

The open loop state optimization problem given an initial state $x_0$ is:
\begin{align}\label{cost_open}
\{ u_k^*\}_{k = 0}^{N - 1} = &\null  \arg \min_{\{u_k\}} \bar{J} (\{x_k\}_{k=0}^N, \{u_k\}_{k=0}^{N-1}), \nonumber \\
&\null x_{k + 1} = f(x_k, u_k, 0), y_k = x_k.
\end{align}

The problem is solved using an NLP solver, where the underlying dynamic models are not needed. For example, open loop optimization using gradient descent  is summarized  as follows.

Denote  the initial guess of the control sequence as $U^{(0)} = \{u_k^{(0)} \}_{k = 0}^{N - 1}$, and  the corresponding  states $\mathcal{X}^{(0)} = \{x_k^{(0)} \}_{k=0}^N$. 

The control policy is updated iteratively via
\begin{align}
U^{(n + 1)} = U^{(n)} - \alpha \nabla_U \bar{J}(\mathcal{X}^{(n)}, U^{(n)}),
\end{align}
until a convergence criterion is met, where $U^{(n)} = \{u_k^{(n)} \}_{k = 0}^{N - 1}$ denotes the control sequence in the $n^{th}$ iteration, $\mathcal{X}^{(n)} = \{x_k^{(n)}\}_{k=0}^N$ denotes the corresponding states, and $\alpha$ is the step size parameter. 

The gradient vector is defined as:
\begin{align}\label{gradv}
\nabla_U \bar{J}(\mathcal{X}^{(n)}, U^{(n)}) = \begin{pmatrix} \frac{\partial \bar{J}}{\partial u_0} & \frac{\partial \bar{J}}{\partial u_1} & \cdots & \frac{\partial \bar{J}}{\partial u_{N - 1}} \end{pmatrix}|_{\mathcal{X}^{(n)}, \{u_k ^{(n)}\}_{k = 0}^{N - 1}},
\end{align}
and without knowing the explicit form of the cost function, each partial derivative with respect to the $i^{th}$ control variable $u_i$ is calculated as follows:
\begin{align}\label{pd}
\frac{\partial \bar{J}}{\partial u_i}|_{\mathcal{X}^{(n)}, U^{(n)}} = \frac{\bar{J}(\mathcal{X}_i^{(n)}, u_0^{(n)},  \cdots, u_i^{(n)} + h, \cdots, u_{N - 1}^{(n)}) 
- \bar{J}(\mathcal{X}^{(n)}, u_0^{(n)},  \cdots, u_i^{(n)}, \cdots, u_{N - 1}^{(n)})}{h},
\end{align}
where $h$ is a small constant perturbation and $\mathcal{X}_i^{(n)}$ denotes the  state corresponding to the control input 
$\{ u_0^{(n)},  \cdots, u_i^{(n)} + h, \cdots, u_{N - 1}^{(n)} \}$, $i = 0, \cdots, N-1$.

The gradient descent procedure is summarized in Algorithm \ref{algo_gradient}.
\begin{algorithm}[!htbp]
\caption{Gradient Descent Algorithm}
\label{algo_gradient}
 \begin{algorithmic}[1]
 \Require{Initial state  $x_0$, cost function $\bar{J}(.)$, initial guess $U^{(0)} = \{u_k^{(0)} \}_{k = 0}^{N - 1}$,  gradient descent design parameters  $\alpha, h, \epsilon$.}
 \Ensure{Optimal control sequence $\{ \bar{u}_k \}_{k = 0}^{N - 1}$, nominal trajectory $\{\bar{x}_k \}_{k  = 0}^N$}
 \State $n = 0$, set $\nabla_U \bar{J}(\mathcal{X}^{(0)}, U^{(0)}) = \epsilon $.
 \While{$\nabla_U \bar{J}(\mathcal{X}^{(n)}, U^{(n)}) \geq \epsilon $}
 \State Evaluate the cost function $\bar{J}(\mathcal{X}^{(n)}, U^{(n)})$.
 \State Perturb each control variable $u_i^{(n)}$ by $h$ and compute the state $\mathcal{X}_i^{(n)}$, $i = 0, \cdots, N - 1$, calculate the gradient vector $\nabla_U \bar{J}(\mathcal{X}^{(n)}, U^{(n)})$. 
  \State Update the control policy 
 $U^{(n + 1)} = U^{(n)} - \alpha \nabla_U \bar{J}(\mathcal{X}^{(n)}, U^{(n)}).$
 \State $n = n + 1$.
 \EndWhile
 \State $\{\bar{u}_k \}_{k = 0}^{N - 1} =  U^{(n)}$, $\{\bar{x}_k \}_{k  = 0}^N = f(x_0, U^{(n)},0)$.
\end{algorithmic}
\end{algorithm}

\section{Time-Varying Eigensystem Realization Algorithm}\label{sec_2}
Consider the linear time-varying (LTV) system:
\begin{align}\label{perturbation system}
& \delta x_{k+ 1} = A_k \delta x_k + B_k \delta u_k + G_k w_k, \nonumber \\
& \delta y_k = C_k \delta x_k + F_k v_k, 
\end{align}
where $\delta x_k = x_k  - \bar{x}_k$ describes the state deviations from the nominal trajectory, $\delta u_k = u_k - \bar{u}_k$ describes the control deviations,  $\delta y_k = y_k - h(\bar{x}_k, 0)$ describes the measurement deviations, and $A_k \in \Re^{n_x \times n_x}, B_k \in \Re^{n_x \times n_u}, C_k \in \Re^{n_y \times n_x}$.

Define the generalized Markov parameters $h_{k, j}$ as:
\begin{align}
h_{k, j} \begin{cases} = C_k A_{k - 1} A_{k - 2} \cdots A_{j+1} B_j, & \text{ if } j < k - 1, \\
= C_k B_{k -1},  & \text{ if } j = k - 1, \\
 = 0, & \text{ if } j > k - 1,
 \end{cases}
\end{align}

and define the generalized Hankel matrix as: 
\begin{align}\label{hankel}
\underbrace{H_k^{(p, q)}}_{pn_y \times qn_u} = \begin{pmatrix}  h_{k, k-1} & h_{k, k-2} & \cdots & h_{k, k - q} \\
h_{k + 1, k - 1} & h_{k + 1, k -2} & \cdots & h_{k + 1, k-q} \\
\vdots & \vdots & \cdots & \vdots \\
h_{k+ p- 1, k -1} & h_{k + p - 1, k -2} & \cdots & h_{k + p - 1, k -q} \end{pmatrix},
\end{align}
where $p$ and $q$ are design parameters could be tuned for best performance, and $pn_y \geq n_r, qn_u \geq n_r$, $n_r$ is the rank of the Hankel matrix. 

Given the generalized Markov parameters, we construct two Hankel matrices $H_{k}^{(p, q)}$ and $ H_{k + 1}^{(p, q)}$, and then solve the singular value decomposition problem: 
\begin{align}\label{svd}
H_k^{(p, q)} = \underbrace{U_k \Sigma_k^{1/2}}_{O_k^{(p)}} \underbrace{ \Sigma_k^{1/2} V_k'}_{R_{k-1}^{(q)}}.
\end{align}
where $(.)'$  denotes the transpose of $(.)$. Denote the rank of the Hankel matrix $H_k^{(p, q)}$ is $n_r$, where $n_r \leq n_x$. Then 
$\Sigma_k \in \mathbb{R}^{n_r \times n_r}$ is the collection of all non-zero singular values,  and $U_k \in \mathbb{R}^{p n_y \times n_r}$, $V_k \in \mathbb{R}^{q n_u \times n_r}$ are the corresponding left and right singular vectors. 

Similarly, $H_{k + 1}^{(p, q)} = O_{k + 1}^{(p)} R_k^{(q)}$.  

The identified system using time-varying ERA is:
\begin{align}\label{ltv_id}
&\underbrace{\hat{A}_k}_{n_r \times n_r} = (O_{k + 1}^{(p) \downarrow})^{+} O_k^{(p)\uparrow} \nonumber \\
&\underbrace{\hat{B}_k}_{n_r \times n_u} = R_k^{(q)}(:, 1: n_u), \nonumber \\
&\underbrace{\hat{C}_k}_{n_y \times n_r} = O_k^{(p)}(1: n_y, :), 
\end{align}
where $(.)^{+}$ denotes the pseudo inverse of $(.)$, $O_{k + 1}^{(p) \downarrow}$ contains the first $(p- 1) n_y$ rows of $O_{k + 1}^{(p)}$, and $O_k^{(p)\uparrow}$ contains the last $(p-1) n_y$ rows of $O_k^{(p)}$. Here, we assume that $n_r$ is constant through the time period of interest, which could also be relaxed. 

Now the problem is how to estimate the generalized Markov parameters. Consider  the input-output map for system (\ref{perturbation system}) with zero noise and  $\delta x_0 = 0$:
\begin{align}\label{input_output}
\delta y_k = \sum_{j = 0}^{k - 1} h_{k, j} \delta u_j.  
\end{align}

We run $M$ simulations and in the $i^{th}$ simulation, choose input sequence $\{ \delta u_{t, (i)} \}_{t = 0}^k$, and collect the output $\delta y_{k, (i)}$. The subscript $(i)$ denotes the experiment number. Then the generalized Markov parameters $\{ h_{k, j} \}_{j = 0}^k $ could be recovered via solving the least squares problem: 
\begin{align}\label{gmarkov}
&\begin{pmatrix} \delta y_{k, (1)} & \delta y_{k, (2)} \cdots & \delta y_{k, (M)} \end{pmatrix} \nonumber \\
& = \begin{pmatrix} 0 & h_{k, k-1} & h_{k, k- 2} & \cdots & h_{k, 0} \end{pmatrix} 
 \begin{pmatrix} \delta u_{k, (1)} & 
\delta u_{k, (2)} & \cdots & \delta u_{k, (M)} \\ \delta u_{k -1, (1)}& \delta u_{k -1, (2)} & \cdots & \delta u_{k - 1, (M)} \\ \vdots & \vdots & &\vdots \\ \delta u_{0, (1)} & \delta u_{0, (2)} & \cdots & \delta u_{0, (M)} \end{pmatrix},
\end{align} 
where $M$ is a design parameter and is chosen such that the least squares solution is possible.

Notice that we cannot perturb the system (\ref{perturbation system}) directly. Instead, we identify the generalized Markov parameters as follows. 

Run M parallel simulations with the noise-free system:
\begin{align}\label{noise_free_input}
&x_{k + 1, (i)} = f(x_{k, (i)}, \bar{u}_k + \delta u_{k, (i)}, 0), \nonumber \\
&y_{k, (i)} = h(x_{k, (i)}, 0),
\end{align}
where $i = 1, 2, \cdots, M$, and therefore,
\begin{align}\label{noise_free_output}
\delta y_{k, (i)} = y_{k, (i)} - h(\bar{x}_k, 0).
\end{align}
where $(\bar{u}_k, \bar{x}_k)$ is the open loop optimal trajectory. Then we solve the same least squares problem with (\ref{gmarkov}).

For simplicity, we assume that the process noise is added through controllers, and the measurement noise is independent of the state and control variables, i.e., $G_k = B_k$, $F_k = I_{n_y \times n_y}$, while the proposed algorithm is extendable to identify $G_k$ and $F_k$.

In general, the identified deviation system is: 
\begin{align}\label{rom}
&\delta a_{k  +1} = \hat{A}_k \delta a_k + \hat{B}_k \delta u_k +  \hat{G}_k w_k, \nonumber \\
& \delta y_k = \hat{C}_k \delta a_k + \hat{F}_k v_k, 
\end{align}
where $\delta a_k \in \Re^{n_r}$ denotes the reduced order deviation states. 

The time-varying ERA used in this paper is summarized in Algorithm \ref{algo_ltv}. 
\begin{algorithm}[!ttbp]
\caption{LTV System Identification}
 \label{algo_ltv}
  \begin{algorithmic}[1]
 \Require{Nominal Trajectory $\{\bar{u}_k\}_{k = 0}^{N - 1}, \{\bar{x}_k \}_{k = 0}^N$, design parameters $M, p, q$}
 \Ensure{$\{ \hat{A}_k, \hat{B}_k, \hat{C}_k \}$}
 \State $k = 0$
 \While{$k \leq N - 1$}
 \State Identify generalized Markov parameters with input and output experimental data using
 (\ref{gmarkov}), (\ref{noise_free_input}) and (\ref{noise_free_output}). 
 \State Construct the generalized Hankel matrices $H_{k}^{(p, q)}$, $H_{k + 1}^{(p, q)}$  using (\ref{hankel}).
 \State Solve the SVD problem, and construct $\{\hat{A}_k, \hat{B}_k, \hat{C}_k \}$ using (\ref{ltv_id}).
 \State $k = k + 1$.
 \EndWhile
\end{algorithmic}
\end{algorithm}